\documentclass[11pt,twoside]{article}

\usepackage{asp2004}
\usepackage{epsf}
\usepackage{psfig}
\usepackage{lscape}

\newcommand{\Halp}{H${\alpha}$}

\newcommand{\ergs}{erg\,s$^{-1}$}

\newcommand{\Msol}{$\rm M_{\odot}$ }

\newcommand{\Lx}{$\rm L_{\rm X}$}

\newcommand{\hd}{{HD\,161103}}
\newcommand{\gcas}{$\gamma$\,Cas}

\usepackage{graphicx}
\begin{document}

\title{X-ray and optical properties of new \gcas-like objects discovered in X-ray surveys}   
\author{Christian Motch$^{1}$, Raimundo Lopes de Oliveira$^{1,2}$, Ignacio Negueruela$^{3}$, Frank Haberl$^{4}$ and Eduardo Janot-Pacheco$^{2}$}

\affil{$^{1}$Observatoire Astronomique, UMR 7550 CNRS, 11 rue de l'Universit\'e, F-67000 Strasbourg, France \\ $^{2}$Instituto de Astronomia, Geof\'{\i}sica e Ci\^encias Atmosf\'ericas, USP, R. do Mat\~ao 1226, 05508-090 S\~ao Paulo, Brazil \\ $^{3}$DFISTS, EPSA, Universidad de Alicante, E-03080 Alicante, Spain \\ $^{4}$Max-Planck-Institut f\"{u}r extraterrestrische Physik, D-85741 Garching, Germany \\ }

\begin{abstract} 

A growing number of early Be stars discovered in X-ray surveys exhibit X-ray luminosities intermediate between those of normal stars and those of most Be/X-ray binaries in quiescence. Their X-ray spectra are also much harder than those of shocked wind OB stars and can be best fitted by a thin thermal plasma with T $\sim$ 10$^{8}$\,K, added to a cooler and much fainter thermal component. An iron line complex including a fluorescence component is detected in many cases. There is no evidence for coherent pulsations in any of these systems but strong variability on time scales as short as 100\,s is usually observed. Large oscillations with quasi-periods of the order of one hour or more are detected in the X-ray light curves of several sources, but have so far failed to prove to be strictly periodic. The optical and X-ray properties of these new objects strikingly resemble those of the so far unique and enigmatic Be star \gcas\ and define a new class of X-ray emitters. We discuss the possible origin of the X-ray emission in the light of the models proposed for \gcas, magnetic disc-star interaction or accretion onto a compact companion object -- neutron star or white dwarf.

\end{abstract}

\keywords{stars: early-type; stars: emission-line, Be; X-rays: stars ; X-rays: binaries}

\section{Introduction}

X-ray surveys are the seed of many important discoveries in the field of high energy astrophysics. Population properties can be studied from the large source samples collected and interesting outliers can be discovered. The {\it Einstein} satellite was instrumental in discovering stellar X-ray emission in most regions of the HR diagram. Early type stars can emit large amounts of X-rays (up to 10$^{33}$ \ergs\ in the 0.2-4.5 keV range for the most extreme cases) with a rather soft thermal spectral energy distribution with kT $\sim$ 0.5\,keV (Cassinelli et al. 1994). X-ray emission is believed to arise in thermalized shocks produced by instabilities in the high velocity wind, which is driven by the strong radiation pressure of the star. Recent high energy resolution X-ray spectra obtained by Chandra and XMM-Newton have confirmed and precised this picture. They indicate that the most active shocks are likely produced in the deep wind and suggest a highly clumped wind structure. 

A number of relatively faint hard X-ray sources have been identified with Be stars in recent X-ray surveys. Their optical properties are those of early type stars with well developed circumstellar discs. Their X-ray properties are however distinct from those of normal OB or Be stars in showing much harder X-ray spectra and slightly enhanced X-ray luminosities. 

HD 110432 is the first Be star found to exhibit these properties. Proposed as the possible counterpart of a faint X-ray source in the HEAO-1 all sky survey the identification was confirmed by Torrejon \& Orr (2001) from BeppoSAX observations. HD 110432 was serendipitously in the field of three XMM-Newton observations (Obsids 0109480101, 0109480201 and 0109480401 for a total exposure of $\sim$ 150\,ksec). Two other objects, HD 161103 and SAO 49725 were also suspected of being low luminosity hard X-ray emitters from the ROSAT all sky survey (Motch et al. 1997) and confirmed by dedicated XMM-Newton observations (Lopes de Oliveira et al. 2005). Finally, the XMM-Newton galactic plane survey conducted by the Survey Science Center yielded at least two more examples of such objects, the Be star SS 397 and another Be star in the 50 Myr old open cluster NGC 6649 (Motch et al. 2005). 

\section{X-ray and optical properties}
\vskip -0.5cm
\begin{table}
\caption{Main properties of currently known \gcas-analogs and of \gcas\ itself.}
\smallskip
\begin{center}
{\small
\begin{tabular}{llcclc}
\tableline
Object      & Spectral   & \Halp\ EW & \Lx\ {\tiny (0.2-12\,keV)}                & kT1  (kT2)  & $\Gamma$   \\ 
            & type       & (\AA)     & {\tiny (10$^{32}$\ergs)}& (keV)       & {\tiny (ph index)}  \\ 
\hline
HD 110432   & B0.5IIIe   & -52       & 3-7                  & $>$ 30 (4.5)  & 1.5         \\ 
HD 161103   & B0.5V-IIIe & -31       & 4-12                 & 8.0 (0.8)   & 1.7         \\ 
SAO 49725   & B0.5V-IIIe & -30       & 4-12                 & 12.8 (0.9)  & 1.5         \\ 
SS 397      & B0.5Ve     & -33       & 2-4                  & $\sim$ 12   & 1.7        \\ 
NGC 6649    & B1-1.5IIIe & -36       & 5                    & $\sim$ 10   & 1.4        \\ 
\gcas       & B0.5IVe    & -26       & 4-11                 & 11.5 (0.2)  & -          \\ 
\hline
\end{tabular}
}
\end{center}
\label{catalogue}
\end{table}
\vskip -0.5cm
		
As shown in Tab. \ref{catalogue}, the overall optical and X-ray properties of the newly discovered low X-ray luminosity and hard X-ray spectra Be stars appear very similar to those of \gcas. A new class of \gcas-like X-ray stars is clearly emerging and in the following we will make no distinction between the \gcas-analogs and \gcas\ itself. It is worth emphasising how similar all these objects are. First, their spectral types fall in a very narrow range very close to the B0.5 type and exhibit almost identical Balmer line equivalent widths. Although the available optical coverage of the new \gcas-analogs is not very large, we did not find any strong variation of the Balmer profile, neither in the total intensity nor in the V/R profile. Second, their X-ray properties, spectral shapes and luminosities also cover a very narrow range of parameter values. The X-ray energy distribution of the faintest sources can be equally well fitted by a power law or by a thin thermal model. For the brightest X-ray sources with the best signal to noise ratios, it seems clear that the power law model does not provide the best fit. A combination of two or more thermal components usually gives smaller residuals. In addition, the thin thermal model provides a natural explanation for the strength of the highly ionized H and He-like Fe\,K lines. A high resolution spectrum of \gcas \ (Smith et al. 2004) confirms this interpretation based on low energy resolution spectra. The X-ray spectra are thus probably dominated by a high temperature component with a kT $\geq$ 8\,keV, while the low temperature thermal component could represent the ``normal'' shocked wind emission. In all cases for which the Fe\,K complex can be observed with enough details, a rather intense low ionization fluorescence line is detected.  

\begin{figure}[t]
\includegraphics[angle=-90,width=0.95\linewidth,bb = 173 45 451 719, clip=true]{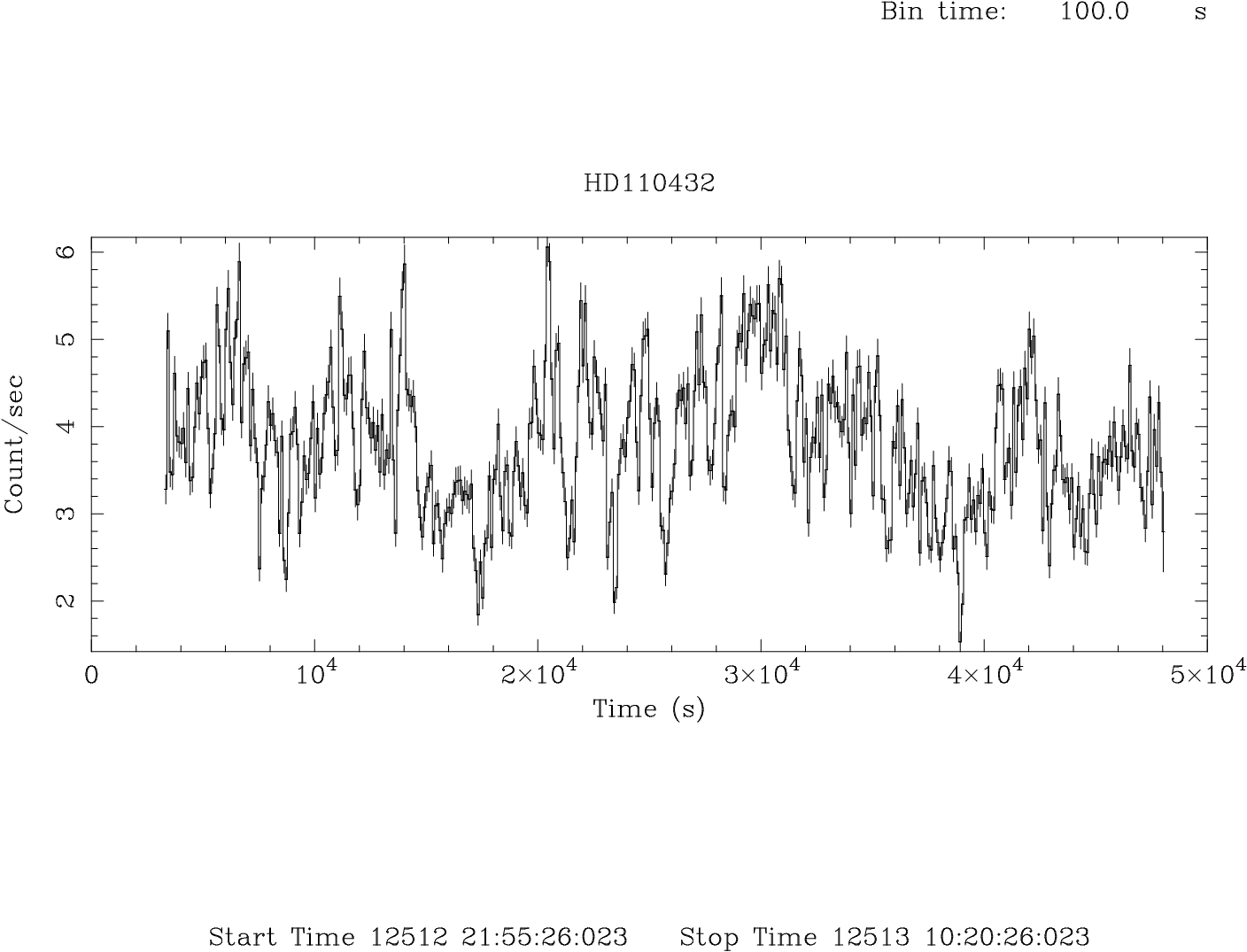}
\includegraphics[angle=-90,width=0.95\linewidth,bb = 173 45 451 719, clip=true]{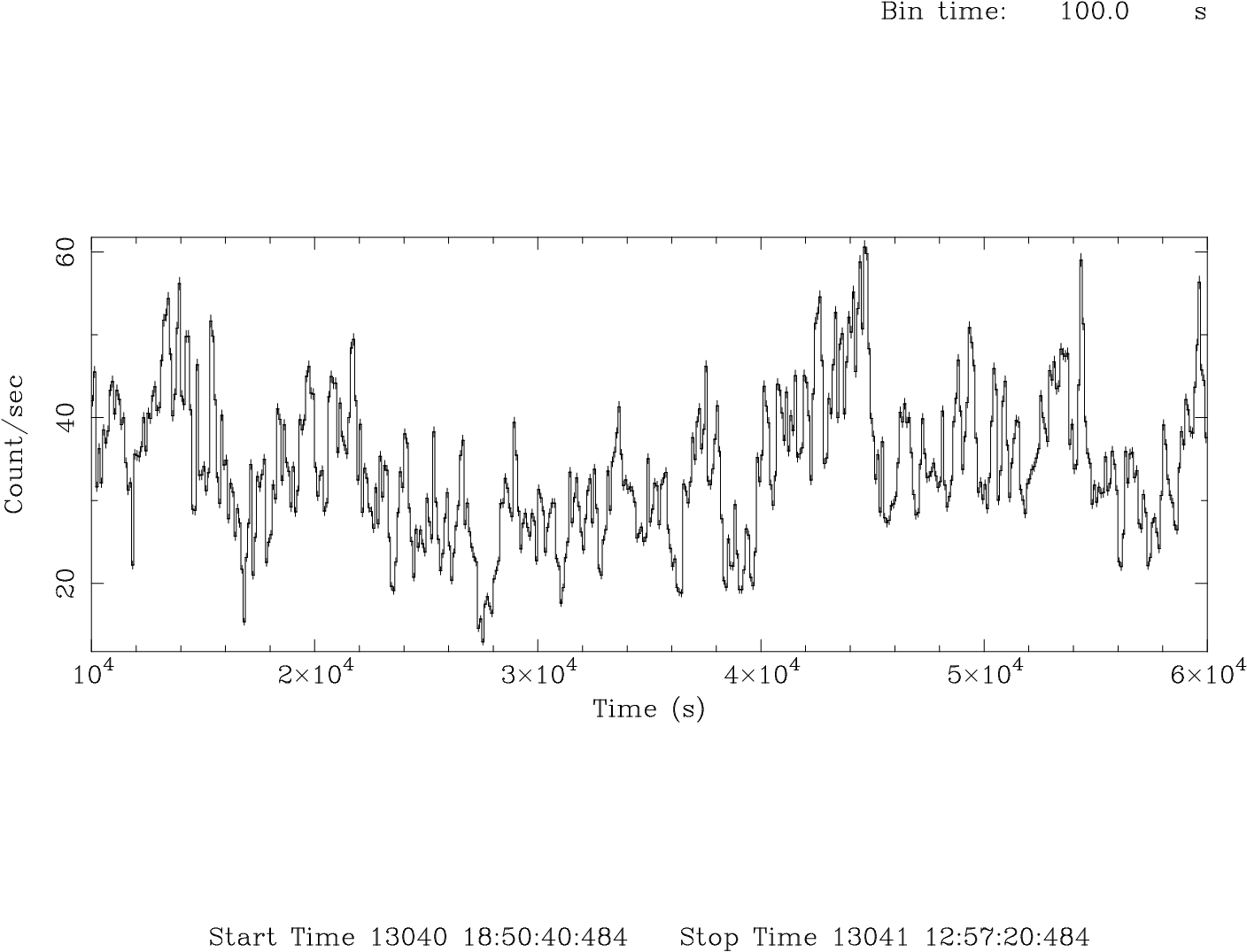}
\caption{XMM-Newton X-ray light curves (0.5-12 keV) of HD 110432 (EPIC pn+MOS, Obsid 0109480201; top) and \gcas\ (EPIC pn, Obsid 0201220101; bottom) binned in 100\,s intervals}
\label{lightcurves}
\end{figure}

The total photoelectric absorption is not very different from that due to the interstellar medium. This indicates that no strong local absorption occurs. However, the possibility remains that part of the high temperature component is significantly more absorbed than other components as is observed in \gcas\ (Smith et al. 2004). 

None of these objects has ever exhibited large X-ray outbursts. Their fluxes vary at most by a factor of a few from one observation to the other. Variability on time scales as short as one hundred seconds or shorter (see Fig. \ref{lightcurves}) is observed from the X-ray brightest members with a 1/f power density spectrum in \gcas\ (Robinson \& Smith 2000) and probably also in HD 110432 (Lopes de Oliveira et al. 2006). Slow variations of relatively large amplitude on hour time scales, sometimes in the form of quasi-periodic modulations are common. For instance, Torrejon \& Orr (2001) report a 233\,mn periodicity in HD 110432 which is not recovered in later XMM-Newton observations (Lopes de Oliveira et al. 2006). A possible 54\,mn period is also seen in \hd\ (Lopes de Oliveira et al. 2005). Several long period modulations (35\,mn, 135\,mn, 145\,mn, 450\,mn, 635\,mn) have been reported by various authors in the X-ray flux of \gcas.

\section{A new class of X-ray emitting Be stars}

\gcas-analogs have X-ray luminosities lying between those of normal OB stars and those of persistent Be/X-ray binaries of low luminosity or of those of most outbursting  Be/X-ray binaries in quiescence. Their soft (0.1-2.4 keV) X-ray luminosities are slightly larger than that of normal B0.5 stars (Cassinelli et al. 1994). However, what distinguishes them most from wind shock emitting stars is their excessively high temperature (kT$\geq$ 8\,keV) compared to kT$\sim$ 0.5 keV typical of normal stars. Some OB stars show weak thermal components of higher temperatures (e.g. Cohen, Cassinelli, \& Waldron  1997) which can at most reach kT $\sim$ 3 keV. Herbig Ae/Be stars display higher temperatures (kT $\sim$ 1 - 5 keV; Hamaguchi, Yamauchi, \& Koyama 2005), which are still significantly lower than those of the \gcas-analogs. Large and fast fluctuations are also not observed in normal OB stars (Cassinelli et al. 1994). 

\begin{figure}[t]
\begin{tabular}{cc}
\includegraphics[angle=-90,width=0.45\linewidth,bb = 81 37 573 710,clip = true]{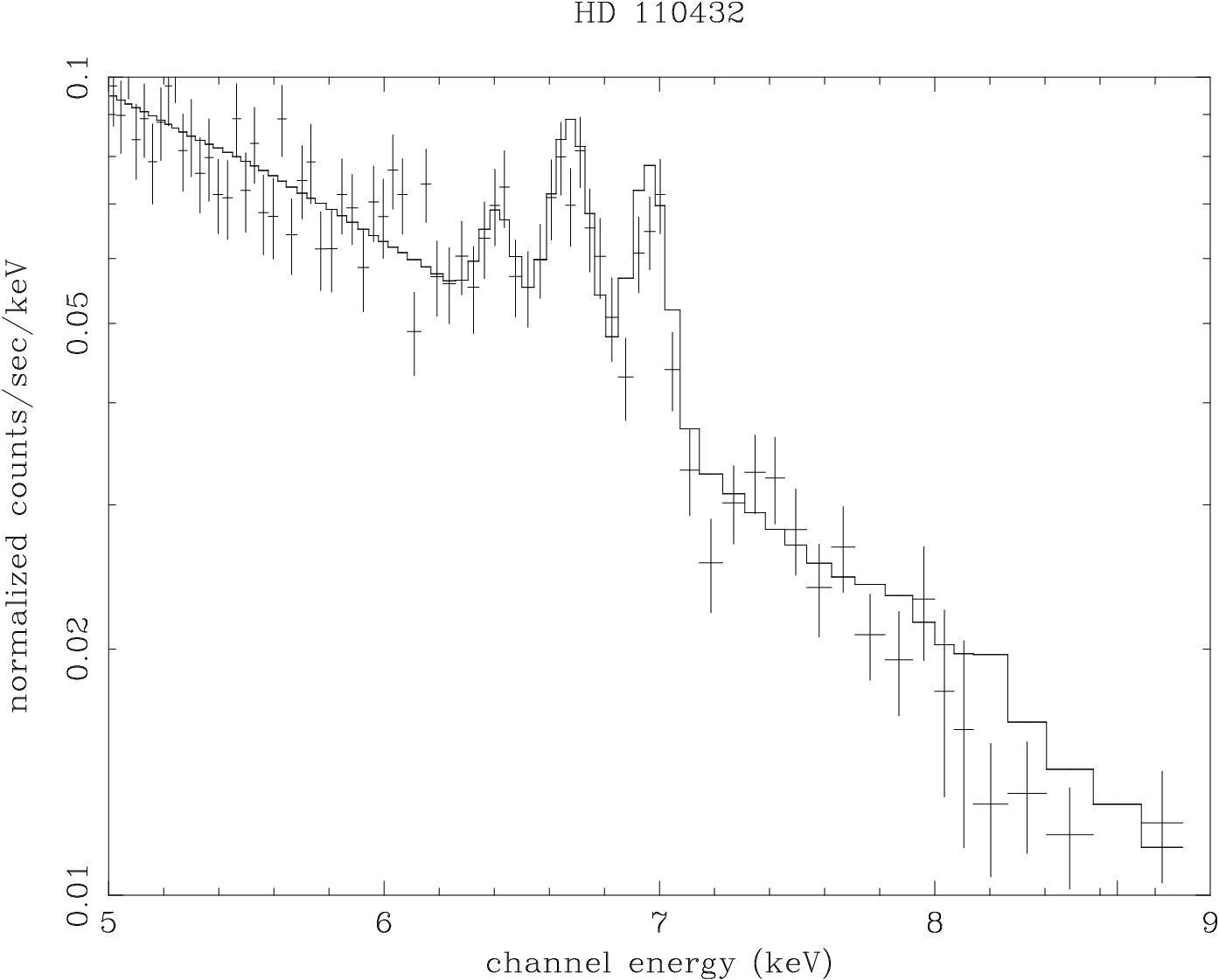} &
\includegraphics[angle=-90,width=0.45\linewidth,bb = 81 37 573 710,clip = true]{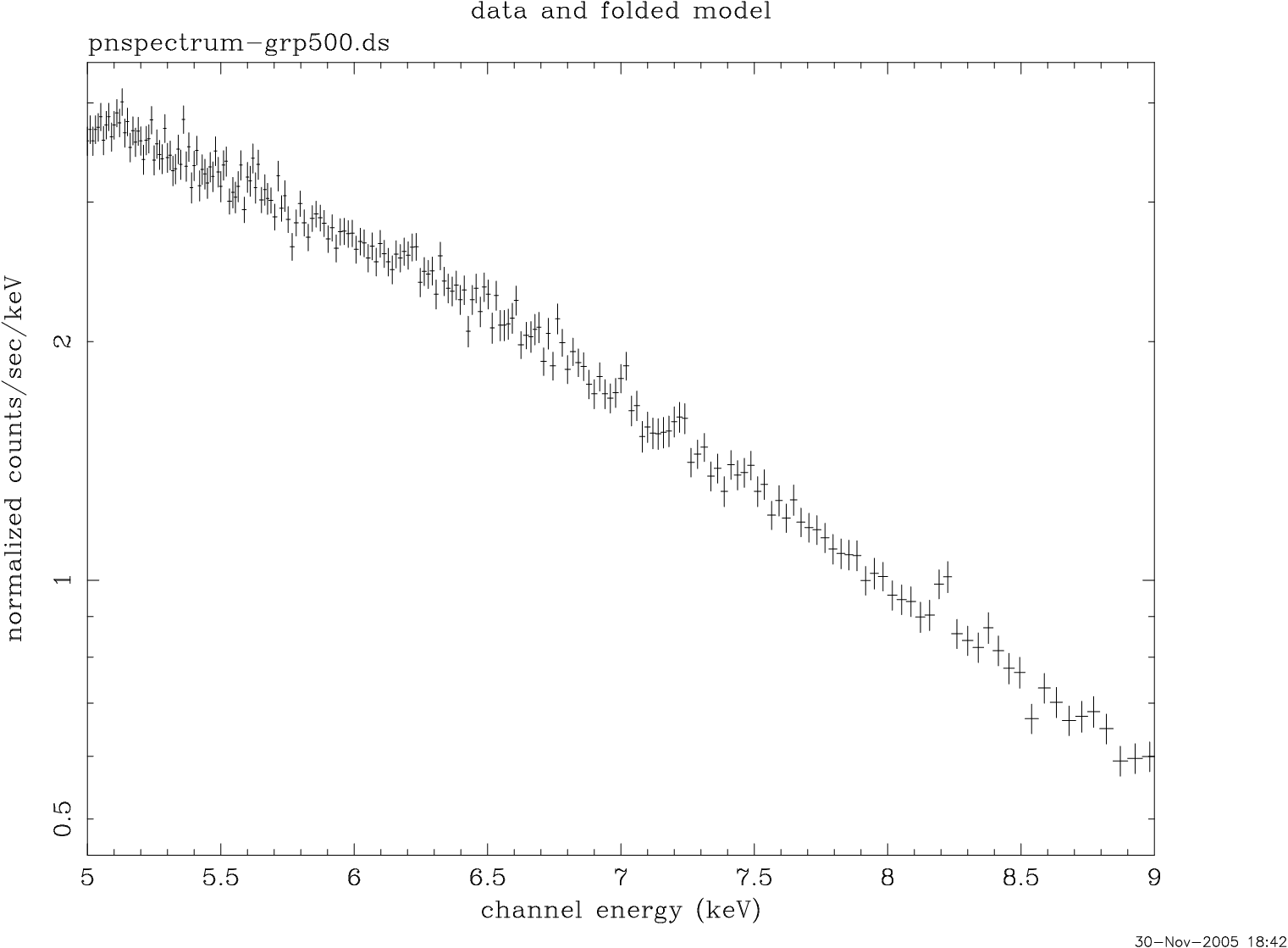} \\
\end{tabular}
\caption{XMM-Newton EPIC pn spectra covering the Fe\,K line complex region. Left; the \gcas-analog HD 110432. Right; the Low \Lx\  persistent Be/X-ray binary X Per (Obsid 0151380101)}
\label{fespec}
\end{figure}

Low X-ray luminosity persistent Be/X-ray stars, the X Per-like objects, have mean X-ray luminosities, log(\Lx) = 34.5, more than an order of magnitude brighter. Another very significant difference is the absence of Fe\,K line emission in these systems (see Fig. \ref{fespec}). This absence is consistent with the essentially non-thermal nature of the X-ray emission from the slowly rotating neutron stars encountered in X Per-like objects. Moreover, the lack of iron fluorescence component probably reflects the low density medium in which the accreting neutron star orbits, namely, the outer rim of the circumstellar Be star discs. These properties are at variance with the likely thermal nature of the X-ray emission in \gcas-analogs and the large equivalent width of the iron fluorescence line which reveals the presence of large amounts of cold material close to the X-ray source. Therefore, although the optical properties of the \gcas-analogs are so far indistinguishable from those of normal Be stars, their X-ray properties are clearly distinct and cannot be easily related to already known classes of X-ray emitters, (apart from \gcas\ itself of course). 

\section{X-ray emission mechanisms}

The origin of the X-ray emission in \gcas\ is a matter of lively debate and still remains very controversial. Two main exclusive explanations have been put forward, accretion onto a compact object, and magnetically heated material at the interface between the photosphere and the disc of the Be star. Unfortunately, so far none of the new \gcas-like stars has brought any definite proof for or against one of the models. The detection of a coherent periodicity that could be interpreted as the spin period of a white dwarf or of a neutron star would be a huge step forward and clearly, more X-ray monitoring is needed for some candidates showing evidences for possibly periodic variations, such as \hd. The systematic detection of companion stars through radial velocity measurements would also be a strong argument in favor of the accreting model.  Proofs or hints indeed exist that two of these Be stars are in binaries. The radial velocity of the wings of the \Halp\ emission line of \gcas\  indicates an orbital period of 204$^{d}$ with $e$ = 0.26 and M$_{2}$ $\sim$ 0.7-1.9 \Msol; Harmanec et al. 2000). Marco, Negueruela, \& Motch (2006) have shown that the star in NGC 6649 is a blue straggler and, as the likely product of massive binary evolution, could harbour a degenerate companion star. 

The weak X-ray luminosities observed in \gcas-analogs are compatible either with accretion at very low rates on a neutron star or even on a black hole, or accretion at a rate similar to that observed in classical Be/X-ray binaries on a relatively weak potential well, such as a white dwarf or the magnetosphere of a fast spinning neutron star in the propeller mode. The strength of the Fe\,K fluorescence line seems to indicate the presence of a large amount of cold material close to the X-ray source and thus probably excludes direct accretion on a neutron star or a black hole, since in order to power no more than a few 10$^{32}$\ergs, the compact object would need to orbit far from the central star in the outskirts of the circumstellar disc. 

Intriguingly, the X-ray spectra of \gcas \ and of its twins strikingly resemble those of cataclysmic variables, in particular those of dwarf novae. Thin thermal components of comparable temperatures are detected in many kinds of CVs. Similar Fe\,K line complexes, including the fluorescence line are seen in some magnetic CVs (Hellier \& Mukai 2004) and many lines observed in the high resolution X-ray spectrum of \gcas\ (Smith et al. 2004) are also found in dwarf novae in quiescence (Pandel et al. 2005). Rapid variability is also observed in many cataclysmic variables, sometimes in the form of shot noise like fluctuations or as a result of blobby accretion onto a magnetic white dwarf. Finally, the strongest argument in favor of an accreting white dwarf might simply be that theories of evolution of massive binaries investigating the binary channel for the production of Be stars predict the existence of a large number of Be + WD systems at a rate which might be up to 7 times that of existing Be + neutron star binaries (e.g. Raguzova 2001). Mass transfer occurring in the late stages of the evolution of the former primary star can lead to binaries consisting of very early B stars with white dwarf companions (Pols et al. 1991). Be + white dwarf binaries are expected to have small eccentricities as a result of the absence of kick at birth and orbital periods of the order of one hundred days. Interestingly, the orbital parameters of \gcas\ with the revised low eccentricity derived by Miroshnichenko et al. (2002) are consistent with those expected for typical Be + WD systems.  The accretion model, however is not free from difficulties. Low eccentricities favor efficient viscous disc truncation (Okazaki \& Negueruela, 2001). It is thus unclear whether the density of disc material available for accretion along the orbit of the white dwarf will be large enough to explain the observed X-ray luminosities. 

A completely different explanation for the X-ray emission of \gcas\ has been proposed by Robinson, Smith, \& Henry (2002) and references therein. The key assumption of the model is that part of the stellar magnetic field is trapped in the inner regions of the ionized circumstellar disc. Field entrainment produces turbulence within the disc and the difference in angular velocities between the disc and the star stresses and shears magnetic lines. Magnetic reconnection leads to the ejection of high velocity plasmoids and generates high energy particles which will be responsible for the hard X-ray emission. This model can adequately explain a number of observational features of \gcas, such as the correlation between the X-ray and optical long term variations and the presence of UV-X-ray correlations on time scale of a few hours, both behaviours finding no straightforward explanation in the accretion model.

The fact that all hard X-ray emitting early type B stars discovered so far have circumstellar discs clearly indicates that the circumstellar disc plays a fundamental role in the generation of the X-rays. This is an obvious statement in the accretion model, but also means that magnetic field connection to the disc is essential in the coronal model. In addition, the very early B types and large Balmer line emission shows that the presence of a well developed decretion disc is a mandatory condition for generating copious hard X-rays. A disc of smaller density may not provide enough material to accrete on a white dwarf to make the small excess of hard X-ray emission detectable. In the magnetic model, this suggests that the density of the disc is also a major ingredient of the model. The presence of a high stellar magnetic field is a strong requirement too. Its origin and geometrical structure is not constrained by the star-disc magnetic model nor by the observations. A fossil origin may explain why not all Be stars with large discs do behave as \gcas. In this context, \gcas-analogs may be related to magnetic early type stars such as the O star $\theta^{1}$ Orionis C (Donati et al. 2002) and Ap and Bp stars in general. Ferrario \& Wickramasinghe (2005) have argued that these objects may well be the progenitors of magnetars.

\section{Conclusions}

In spite of being among the first X-ray sources discovered, the mechanism explaining the X-ray emission of \gcas\ remains uncertain. The discovery of many more twins of this so far unique case will certainly revivify the interest in studying the puzzling origin of this very hard X-ray emission. In addition, these stars could significantly contribute to the population of galactic hard X-ray emitters since they account for nearly half of the Be/X-ray candidates found in the XMM-Newton SSC survey of the galactic plane (Motch et al. 2005). Members of this new class display very different X-ray properties from those of classical Be/X-ray binaries and ``normal'' OB stars. The narrowness of the observed range of X-ray and optical characteristics is intriguing and suggests that rather specific conditions must be met to generate the hard X-ray luminosity observed. Whatever is the true X-ray emitting mechanism, the discovery of Be + WD binaries or of a population of magnetically active early type stars is an exciting prospect.

\acknowledgements 
 RLO  acknowledges financial support from Brazilian agencies FAPESP (grant 03/06861-6) and CAPES (grant
BEX0784/04-4), and Observatoire de Strasbourg.
IN is a researcher of the
 programme {\em Ram\'on y Cajal}, funded by the Spanish Ministerio de
 Ciencia y Tecnolog\'{\i}a (currently Ministerio de Educaci\'on y
 Ciencia) and the University of Alicante, with partial
 support from the Generalitat Valenciana and the European Regional
 Development Fund (ERDF/FEDER).
 This research is partially supported by the MCyT (currently MEC) under
 grant AYA2002-00814. This work is based on observations obtained with XMM-Newton, an ESA science mission with instruments and contributions directly funded by ESA Member States and NASA.


\end{document}